\pdfoutput=1
\documentclass[unnumsec,webpdf,contemporary,large]{oup-authoring-template}

\graphicspath{{./}{Fig/}}

\usepackage{amsmath,amssymb,amsfonts}
\usepackage{soul}
\usepackage{xcolor}
\usepackage{makecell}
\usepackage{amsmath}

\setcellgapes{0pt}
\makegapedcells
\sethlcolor{yellow}

\usepackage{comment}
\usepackage{graphicx}
\usepackage{subfig}

\usepackage{caption}

\begin{document}

\journaltitle{Journal Title Here}
\DOI{DOI added during production}
\copyrightyear{YEAR}
\pubyear{YEAR}
\vol{XX}
\issue{x}
\access{Published: Date added during production}
\appnotes{Paper}
\firstpage{1}

\title[mUTR-Construct: A Conditional Structure-Aware Generative Transformer for Multi-Objective Design of Modified RNA 5' UTRs]{mUTR-Construct: A Conditional Structure-Aware Generative Transformer for Multi-Objective Design of m1$\Psi$-Modified RNA 5' UTRs}

\author[1]{Narges Zarnaghinaghsh}
\author[2]{Ahmadreza Mofayezi}
\author[1,3]{Byung-Jun Yoon}

\address[1]{Texas A\&M University, Department of Electrical and Computer Engineering, College Station, TX 77843, USA.}
\address[2]{The Netherlands Cancer Institute, 1066 CX Amsterdam, Netherlands.}
\address[3]{Brookhaven National Laboratory, Computing and Data Sciences, Upton, NY 11973, USA.}

\corresp[$\ast$]{Corresponding author. \href{mailto:email-id@example.com}{nzarnaghi@tamu.edu}}

\received{Date}{0}{Year}
\revised{Date}{0}{Year}
\accepted{Date}{0}{Year}

\abstract{The 5' untranslated region is a major determinant of translation initiation, and its effect becomes especially important in modified mRNA sequences, where start-codon context, cap-proximal secondary structure, upstream AUGs and upstream open reading frames, and nucleotide chemistry can alter ribosome scanning and initiation recruitment, scanning, and decoding in sequence-dependent ways. Recent computational studies have moved the field from prediction toward design, including massively trained predictive models such as Smart5UTR for m1$\Psi$-modified mRNA, broader 5' UTR generation and optimization frameworks such as UTRGAN and UTailoR, and structure-guided RNA design systems such as RhoDesign. Here, we describe a conditional generative framework for 50-nt modified-RNA 5' UTR design that optionally conditions on ribosome load, GC content, minimum free energy, and target secondary structure. The implementation uses a Transformer-based generator followed by sequence ranking and local refinement with a Smart5UTR-derived ribosome-load oracle and ViennaRNA-based folding metrics, including support for modified-base folding parameters. Across multiple simulation scenarios and experimental settings, different combinations of RL, GC, MFE, and structural constraints produced distinct performance tradeoffs, enabling ablation-based identification of the best-performing formulation. In a benchmark focused on target-structure design, the proposed framework achieved a higher success rate than the compared RhoDesign baseline in the modified-RNA setting studied here; a cautious interpretation is that the present pipeline explicitly evaluates modified-RNA folding behavior during search and ranking, whereas RhoDesign was developed for canonical RNA inverse folding. \newline \textit{Code availability:} The source code of RGVAE and the data used in this study can be found at: {https://github.com/nzarnaghinaghsh/mUTR-Construct}.}


\maketitle

\section{Introduction}
The 5' untranslated region (5' UTR) is a non-coding sequence located before the protein-coding region of an mRNA. Although it is not translated into protein \citep{wieder2024differences}, its sequence can strongly affect the amount of protein produced from the mRNA \citep{chu20245, dvir2013deciphering, cuperus2017deep}. For computational sequence design, the 5' UTR is therefore an important input region because small sequence changes can produce measurable changes in protein expression \citep{dvir2013deciphering, cuperus2017deep}.

Large-scale experimental datasets have made it possible to model the relationship between a 5' UTR sequence and translation-related measurements \citep{karollus2021predicting}. Cuperus et al. constructed and measured approximately 500,000 random 50-nt yeast 5' UTRs and trained a convolutional neural network to predict protein expression \citep{cuperus2017deep}. Sample et al. combined polysome profiling with a library of 280,000 randomized human 5' UTRs and developed a model that predicts mean ribosome load (MRL) \citep{sample2019human}. Their work also showed that a prediction model can be combined with a search algorithm to design sequences for specified ribosome-loading levels \citep{sample2019human}. These studies established a practical machine-learning formulation in which a nucleotide sequence is used as input and a translation-related quantity is used as the prediction target.

Subsequent models expanded the range of computational methods used for 5' UTR analysis. UTR-LM applied language-model pretraining to endogenous 5' UTR sequences and added supervised information on secondary structure and minimum free energy (MFE) \citep{chu20245}. It was evaluated on MRL, translation efficiency, and mRNA-expression prediction tasks and was also used to design new sequences \citep{chu20245}. Transfer-learning approaches have been investigated to improve prediction across experimental contexts \citep{gilliot2024transfer}. More recent models have also combined prediction with sequence optimization for mRNA-delivered gene editing and other therapeutic applications \citep{castillo2024optimizing}.

Modified nucleotides create an additional design requirement. N1-methylpseudouridine (m1$\Psi$) is widely used in synthetic mRNA, but its effect is not identical for every sequence \citep{tang2024novel, svitkin2017n1, lewis2025quantitative, rozman2026n}. Smart5UTR was developed specifically to predict and design 50-nt 5' UTRs for m1$\Psi$-modified mRNA \citep{tang2024novel}. Experimental and computational studies have reported that the effect of m1$\Psi$ depends on sequence context and can alter ribosome loading or translation dynamics \citep{svitkin2017n1, lewis2025quantitative, rozman2026n}. These findings motivate models that are trained and evaluated specifically for modified RNA rather than assuming that models developed for canonical RNA transfer without modification-specific analysis.

Research has increasingly moved from prediction to generative design. Cao et al. used naturally occurring 5' UTRs and an in-silico genetic algorithm to generate synthetic sequences with increased protein expression \citep{cao2021high}. UTRGAN introduced a generative-adversarial-network pipeline for generating and optimizing variable-length human 5' UTRs \citep{barazandeh2025utrgan}. UTailoR combined a discriminative translation-efficiency predictor with a generative model that modifies an input 5' UTR while retaining sequence similarity \citep{liu2025enhancing}. These methods illustrate a general design workflow: train a predictive or generative model from sequence–measurement pairs, propose candidate sequences, evaluate them with one or more objectives, and retain improved candidates.

Structure-aware design has also been investigated in the literature. Structure-aware design is important because structure plays a crucial role in the function of RNA sequences. ViennaRNA provides widely used algorithms for RNA secondary-structure and MFE calculations \citep{lorenz2011viennarna}. RhoDesign demonstrated a structure-to-sequence deep-learning approach for RNA aptamer generation using predicted three-dimensional structures and additional structural inputs \citep{wong2024deep}. These developments support the inclusion of the target structure as explicit computational objectives rather than evaluating generated sequences only through predicted translation. 

Multi-objective design introduces a further challenge because MRL, GC content, MFE, structure similarity, and motif requirements may not improve simultaneously. Transformer architectures provide a complementary sequence-modeling component by producing nucleotide probability distributions conditioned on the available sequence and target information \citep{vaswani2017attention}. 

In this study, we develop a conditional framework for designing 50-nt 5' UTR sequences for m1$\Psi$-modified RNA. The model uses the sequence together with ribosome load (RL), GC content, MFE, and optional secondary structure during training. During design, the conditional Transformer samples candidate sequences from nucleotide probability distributions rather than applying deterministic highest-probability decoding. Candidate sequences are subsequently evaluated using Smart5UTR-predicted RL, GC content, ViennaRNA-derived MFE and structure, and optional motif constraints. Two refinement strategies are investigated: a hybrid beam-based local optimizer that ranks complete sequences using a combined score, and a Pareto-based population optimizer that uses non-dominated sorting and generator-guided mutations. Ablation experiments evaluate different combinations of RL, GC, MFE, and structural conditioning, while comparisons with Smart5UTR and RhoDesign assess translation-target and structure-design performance.

\section{Methods}
\begin{figure*}
\begin{center}
\includegraphics[width=6.8 in]{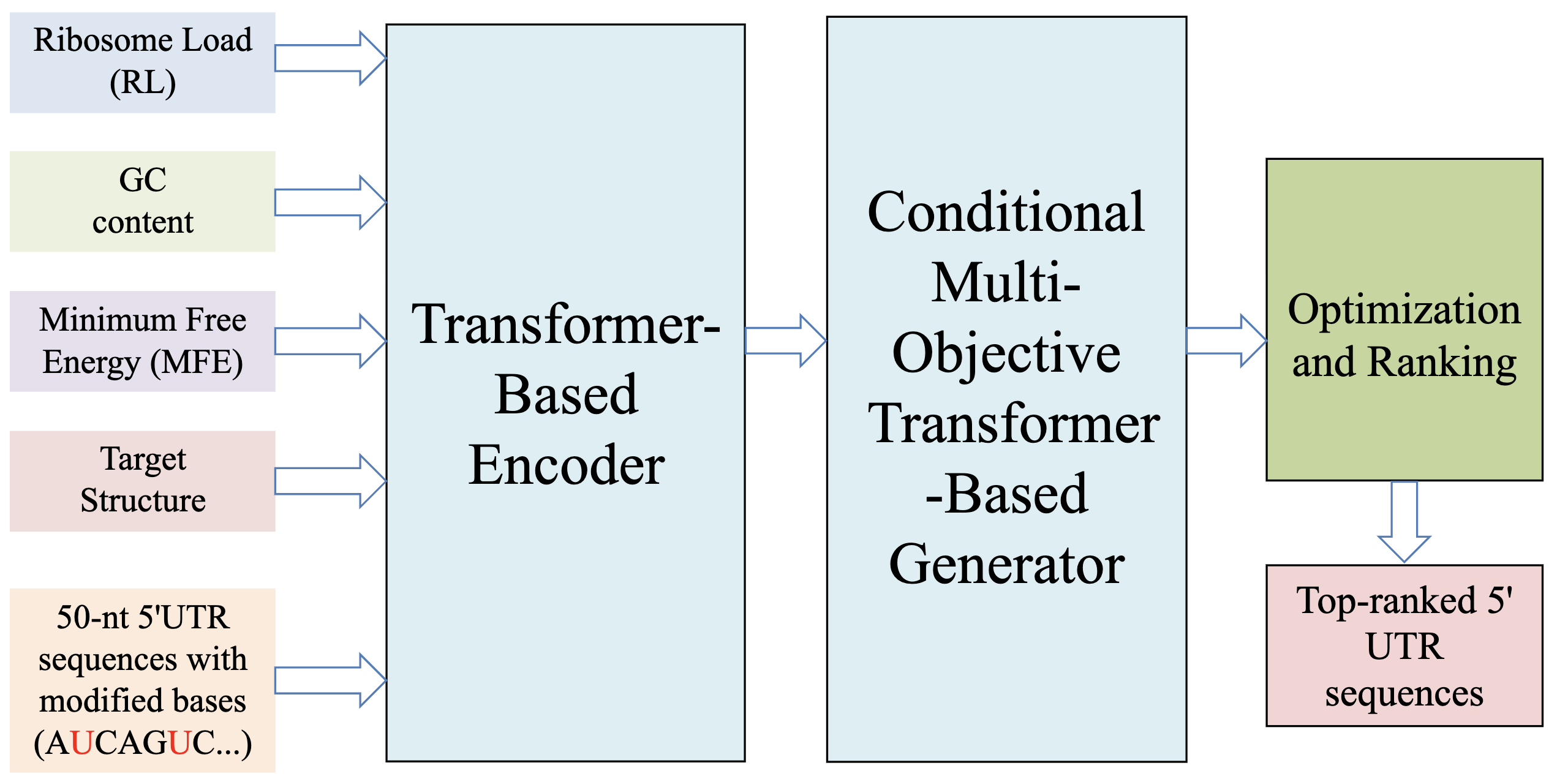}
\end{center}
\caption{The diagram of the mUTR-Construct model.\label{model}}
\end{figure*}

The proposed pipeline defines a conditional structure-aware generative Transformer for multi-objective design of m1$\Psi$-modified 50-nt RNA 5' UTR sequences which is shown if Figure \ref{model}. The core generator is a Transformer-style encoder model with nucleotide token embeddings, positional encoding, and a condition embedder that projects scalar targets for ribosome load, GC content, MFE, and the structure into the latent space. After doing hyper-parameter optimization and five-fold cross validation, the default generator hyperparameters in the code are sequence length 50, hidden dimension 64, 4 attention heads, 2 encoder layers, feed-forward dimension 256, dropout 0.1, learning rate 1e-4, batch size 32, and 5 training epochs. The model is trained to reconstruct sequences under the selected optional conditions for target RL, GC content, MFE, structure, required motifs, and forbidden motifs. 

Ribosome-load scoring is handled by a Smart5UTR scorer wrapper that loads the Smart5UTR model and scaler, encodes sequences in one-hot form, and predicts scalar RL values for candidate 5' UTRs. The same code computes GC fraction directly from sequence and computes MFE plus folded structure with ViennaRNA. The composite design score is defined as the sum of a negative RL absolute error term, a negative GC absolute error term, a negative MFE absolute error term, and a positive structure-match term, with optional motif contributions. In other words, the design objective can be understood as learning and then searching over an approximation to

\begin{equation}
\begin{aligned}
\mathrm{Score} ={}&
-|RL-\operatorname{target}(RL)|
-|GC-\operatorname{target}(GC)| \\
&-|MFE-\operatorname{target}(MFE)| \\
&+\operatorname{Match}\bigl(\mathrm{Structure},
\operatorname{target}(\mathrm{Structure})\bigr) \\
&+\operatorname{Match}(\mathrm{Motif\ Term})
\end{aligned}
\end{equation}

This matches the user’s project description: sequence generation is not driven by RL alone, but by joint control of physicochemical and structural features.  

\subsection{Hybrid Sequence Generation and Beam-Based Optimization}
In the hybrid method, for a specified set of design objectives, including target RL, GC content, minimum free energy (MFE), and optional secondary-structure constraints, the proposed framework first generates an initial pool of candidate 5' UTR sequences using a conditional Transformer-based generator. Unlike deterministic decoding approaches that select the nucleotide with the highest predicted probability at each position, the proposed method employs probabilistic sampling from the predicted nucleotide distribution. Specifically, at each sequence position, the Transformer outputs a probability distribution over the four RNA nucleotides (A, C, G, and U). Candidate nucleotides are then sampled according to these probabilities, allowing multiple distinct sequences to be generated while preserving the learned sequence statistics of the training data. 

For a target design condition, the generator produces (N=8) complete candidate sequences. Each candidate sequence is evaluated using a multi-objective scoring function that incorporates the optional desired Smart5UTR-predicted ribosome load, GC-content deviation from the target value, MFE deviation from the target value, secondary-structure similarity, and motif constraints based on the target objectives. The generated candidates are ranked according to their overall objective score, and the top (k) candidates are retained as the initial beam population. To further improve objective satisfaction, a beam-based local optimization procedure is applied. During each optimization iteration, every sequence within the beam generates a set of neighboring candidate solutions through nucleotide mutations. Specifically, 12 mutant sequences are produced from each beam sequence by randomly modifying multiple nucleotide positions. The number of mutated positions is adaptively reduced throughout the search process, allowing broad exploration during early iterations and fine-grained refinement during later iterations. 

All mutant sequences are evaluated using the same multi-objective scoring function employed during candidate ranking. The scoring function combines the ribosome-load objective, GC-content objective, MFE objective, secondary-structure objective, and motif constraints into a unified optimization score. Following evaluation, all candidate sequences generated during the iteration are ranked, and the highest-scoring (k) sequences are retained for the next optimization cycle. This beam-maintenance strategy enables simultaneous exploration of multiple promising regions of sequence space while avoiding premature convergence to a single local optimum. 

For each target condition, the conditional Transformer initially generates eight candidate 5' UTR sequences through probabilistic sampling. These sequences constitute the initial search population. The subsequent optimization stage operates exclusively on mutated variants of the retained beam sequences and does not invoke additional Transformer-based sequence generation. 

After initial conditional sampling, the top-ranked generated sequences were refined using a beam-based local search. The optimization procedure is repeated for 60 iterations.  At each iteration, multiple mutated variants were generated from every beam sequence, evaluated using the multi-objective score, and the best-scoring candidates were retained. After the iterative search, a final single-base refinement step was applied to the top beam sequences. In this step, a subset of nucleotide positions was selected randomly, and all three alternative nucleotide substitutions were evaluated at each selected position. The sequence achieving the highest overall objective score after refinement is reported as the final optimized 5' UTR design. 

The overall workflow consists of: (1) conditional sequence generation through probabilistic sampling from Transformer-predicted nucleotide distributions, (2) selection of the top-scoring beam candidates, (3) iterative mutation-based beam optimization using 12 neighboring candidates per beam sequence, (4) multi-objective evaluation based on RL, GC content, MFE, structure, and motif constraints, and (5) final local refinement and selection of the highest-scoring sequence. 

\subsection{Pareto-Based Multi-Objective Optimization}
To provide a baseline multi-objective optimization strategy, a Pareto-based optimization framework was implemented and compared with the proposed hybrid generation approach. Unlike the hybrid method, which combines all biological requirements into a unified objective function, the Pareto method treats each design criterion as an independent optimization objective and seeks solutions that achieve the best trade-offs among competing requirements. 

For each generated RNA sequence, multiple biological properties are evaluated independently, including the Smart5UTR-predicted ribosome load (RL), GC content, minimum free energy (MFE), secondary-structure similarity, and optional motif constraints. Let (x) denote a candidate sequence and let 

$\mathbf{f}(x)=\left[f-{RL}(x), f-{GC}(x), f-{MFE}(x), f-{Struct}(x)\right] $

represent the corresponding objective vector. 

The RL objective is defined as the absolute deviation between the predicted ribosome load and the target value, while the GC-content and MFE objectives measure the differences between the generated and desired biological characteristics. When structural constraints are imposed, a secondary-structure similarity score based on dot-bracket matching is also incorporated. Motif constraints are handled independently through the presence or absence of required and forbidden sequence patterns. 

The optimization procedure begins with the generation of an initial population of candidate RNA sequences using the conditional generator. Each sequence is subsequently evaluated according to all objective functions. Rather than combining these objectives into a single scalar score, Pareto dominance relationships are established among candidate solutions. 

A candidate sequence $x-i$ is said to dominate another sequence $x-j$ if it performs at least as well in every objective and strictly better in at least one objective. Formally, $x-i$ dominates $x-j$ if

$f-k(x-i) \geq f-k(x-j), \quad \forall k, $

and

$f-k(x-i) > f-k(x-j)$

for at least one objective $k$. The set of non-dominated solutions constitutes the Pareto front and represents the best achievable trade-offs among the biological requirements. 

A sequence is in the first Pareto front if no other candidate sequence is better than it across all objectives. For example, one sequence may have better RL matching but worse structure matching, while another may have better structure matching but worse GC or MFE. If neither sequence is strictly better in all objectives, both can remain in the first Pareto front. Following non-dominated sorting, sequences belonging to the first Pareto front are retained as candidate designs. If additional optimization iterations are performed, mutation operators are applied to generate neighboring solutions, and the Pareto-ranking procedure is repeated to progressively improve the population while preserving objective diversity. The final output consists of the highest-quality non-dominated solutions that simultaneously satisfy the competing constraints without requiring manually selected weighting coefficients. 

The principal advantage of the Pareto framework is that it avoids imposing subjective priorities among RL, GC content, MFE, and structural objectives. Instead, it provides a set of biologically meaningful trade-off solutions, allowing different design criteria to be satisfied to varying degrees. This property makes Pareto optimization a useful baseline for evaluating whether the proposed hybrid framework can achieve superior objective satisfaction while maintaining sequence diversity and structural feasibility. 

In the Pareto method, refinement was performed through a population-based Pareto search rather than through the final single-base beam refinement used in the hybrid method. An initial population was sampled from the conditional generator, evaluated across the independent objectives, and ranked by Pareto dominance. During Pareto optimization, the generator was used to identify sequence positions exhibiting the greatest prediction uncertainty. Shannon entropy was computed from the nucleotide probability distribution at each position, and positions with the highest entropy were preferentially selected for mutation. For each selected position, the current nucleotide was down-weighted and a replacement nucleotide was sampled from the renormalized generator distribution. Consequently, mutations were guided by the learned sequence distribution rather than being applied uniformly at random. Offspring were evaluated and merged with the current population, and the next generation was selected using non-dominated sorting and crowding distance. The final design was selected from the first Pareto front according to the total objective score. 

The overall Pareto optimization workflow consists of: (1) generation of candidate RNA sequences, (2) independent evaluation of RL, GC content, MFE, structure, and motif objectives, (3) Pareto dominance analysis and non-dominated sorting, (4) preservation of Pareto-optimal solutions, (5) mutation-based generation of neighboring candidates, and (6) selection of the final non-dominated RNA designs.

\section{Results and Discussion}

Three main conclusions emerge from the experiments. First, all learned generator variants dramatically outperform the reported Smart5UTR search baseline in the ablation study. Second, the selected structure-aware hybrid model is robust across a wide range of scenario families, including RL-only, multi-objective RL+GC and RL+MFE settings, structure-constrained design, motif-constrained design, and several “real prior” tests such as Kozak, top-like, and accessibility matching. Third, the proposed method exceeds the RhoDesign baseline in secondary-structure fidelity in the benchmark examined here, though that comparison must be interpreted cautiously because the two runs do not share the same target structure or objective set.

All experiments were conducted on an EGFP $m1\Psi$-modified dataset and used modified-base folding parameters via a ViennaRNA JSON file, with the modified base encoded as symbol 1 and mapped to U for folding when needed. The scripts show a  canonical target configuration of RL $ = 7.5$, MFE $ = - 8.0$, GC $ = 0.5$, and a fixed 50-nt target structure for the ablation runs. The code also allows GPU training and CPU generation, which is consistent with the supplied SLURM workflow. 

In the following subsections, the ablation study results, the results of different scenarios, the results of the specific applications, and the benchmark results against RhoDesign model are provided.

\subsection{Ablation Study}

For designing the architecture of the model, an ablation study is performed to evaluate the effectiveness of incorporating different features into the design. The ablation runner explicitly defines four learned variants plus a baseline. RL-GC-MFE uses RL, GC, and MFE; RL-only removes GC and MFE and effectively becomes RL-only; RL-GC-MFE-struct uses RL, GC, MFE, and structure; and RL-structure uses RL plus structure. A separate smart5utr-baseline calls external optimization routines around the Smart5UTR predictor. The same optimizer is then used for generator-based methods, making the ablation comparison interpretable as a test of which conditioning signals add value under a shared search procedure. 

The ablation pattern is revealed in Table \ref{Abl}. The fully structure-aware model (RL-GC-MFE-struct) achieved the best overall balance, with the highest mean total score, the strongest structural agreement, and the lowest MFE mean absolute error (MAE). The RL-structure model was almost as strong overall and unexpectedly produced the best RL and GC errors, suggesting that explicit structural conditioning may act as a compact surrogate for constraints otherwise only partially captured by GC and MFE targets. By contrast, the Smart5UTR baseline fell far behind every multi-objective quality metric.

\begin{table*}[h]

\centering
\small
\sffamily


{\centering
\caption{Ablation study summary (n = 200 designs per method).}\label{Abl}
\par}

\begin{tabular}{|c|c|c|c|c|c|c|c|c|}
\hline
 Variant&Conditioning signals&Mean optimized RL&RL MAE&GC MAE&MFE MAE&Mean structure match&Success rate&Mean runtime\\
\hline
RL-GC-MFE&RL+GC + MFE&7.4928&0.0087&0.0204&0.0040&0.6344&0.990&9.77 s\\

\hline
RL-only &RL only&7.4946&0.0068&0.0151&0.0097&0.6329&0.995&9.74 s\\

\hline
RL-GC-MFE-struct &RL+GC+MFE+structure&7.4954&0.0069&0.0114&\textbf{0.0030}&\textbf{0.6531}&0.995&9.98 s\\

\hline
RL-structure&RL+structure&7.4970&\textbf{0.0060}&\textbf{0.0071}&0.0047&0.6491&0.995&10.51 s \\

\hline
smart5utr-baseline&Baseline search&7.0923&0.4105&0.0671&4.0038&0.6078&0.155&25.74 s \\

\hline
\end{tabular}\\[10pt]
\end{table*}

In the following sections, we used the model RL-GC-MFE-struct as mUTR-Construct model for the evaluation.

\subsection{Optimization results}

In this section, the optimization results for different scenarios are provided. Since different conditions affect on each other, we implemented different scenarios with different combinations of constraints for evaluating the performance of the proposed method.  The total number of the generated sequences for each of the optimization scenarios is 500. The structure-aware mUTR-Construct model is consistently the most accurate method overall, the Smart5utr baseline is often competitive on pure RL targeting, and Pareto method offers a speed–accuracy trade-off that is attractive computationally but usually unacceptable when tight objective matching is required. 

The constraints we evaluated for optimizing the generated sequences are as follows:
\begin{itemize}
    \item \textbf{ribosome load (RL) Constraint:} The aim is to reach a target RL value. 
    \item \textbf{Minimum Free Energy (MFE) Constraint:} The aim is to reach a specific MFE Value. Generally, RNA sequences with thermodynamically stable RNA secondary structures have lower free-energy \citep{zarnaghinaghsh2025efficient}.
    \item \textbf{GC-content Constraint:} The percentage of the C or G nucleotides in an RNA sequence is the GC-content. The aim of the optimization is to reach a target GC-content.
    \item \textbf{Secondary structure constraint (CT):} The goal is generating sequences with a particular target secondary structure.
    \item \textbf{Mandatory motifs constraint (CM):} The goal is to generate sequences that contain a specific group \textit{M} of sequences at any position at least one time.
    \item \textbf{Forbidden motifs constraint (CF):} The aim is to generate sequences such that a predefined group \textit{F} of forbidden sequences should not exist in any of them.
\end{itemize}

\subsubsection{RL-only Optimization}

In this subsection, the optimization results for only the RL feature are provided. Table \ref{RL-opt} shows the results of the hybrid optimization, the Pareto optimization, and the Smart5UTR baseline. For target RL values of 8, 7, 5, and 3, the hybrid method achieved mean RL absolute errors of 0.00005, 0.00003, 0.00007, and 0.00007, respectively, with 100\% success in every case. The Smart5UTR baseline was similarly near-exact, with mean RL absolute errors between 0.00008 and 0.00019 and 100\% success across the board. The Pareto method was accurate at low and moderate RL targets, but its performance collapsed at the hardest high-expression target of RL = 8, where mean RL absolute error rose to 0.8874 and success fell to 0.4\%. This subsection can therefore state that the proposed model preserves exact RL controllability while remaining more stable than Pareto method at the upper end of the target range. In addition, the mean of the total score for the hybrid method is higher than all of the other methods for all of the target values, but the mean of the runtime is also higher than the other methods. 

\begin{table*}[h]
\small\sf\centering
\begin{center}
\centering
\captionsetup{justification=centering}
\caption{The optimization results for optimizing only target RL values. \label{RL-opt}}

\begin{tabular}{|c|c|c|c|c|c|c|c|}
\hline
Target RL&Method&Mean RL&RL MAE&Rmse RL&Success RL tolerance rate&Mean total score&Mean runtime (sec) \\

\hline
8&hybrid&8.000000864&5.03E-05&7.43E-05&1&-2.21E-08&26.72199192\\

\hline
8&Pareto&7.112616619&0.887416519&0.919815435&0.004&-3.38424174&2.088339616\\

\hline
8&smart5utr&7.999991817&0.000134641&0.00020967&1&-1.76E-07&8.625981972\\

\hline
7&hybrid&7.000000072&2.91E-05&4.24E-05&1&-7.19E-09&27.99102738\\

\hline
7&Pareto&6.9482514&0.055644924&0.107679029&0.784&-0.046379093&2.049192488 \\

\hline
7&smart5utr&7.000001014&7.71E-05&0.000117971&1&-5.57E-08&8.909683352\\

\hline
5&hybrid&4.999995667&6.64E-05&9.40E-05&1&-3.53E-08&30.46960977 \\

\hline
5&pareto&5.002558353&0.008920565&0.014278923&0.998&-0.000815551&2.143904806\\

\hline
5&smart5utr&4.999994731&0.000164638&0.000233293&1&-2.18E-07&9.420545704\\

\hline
3&hybrid&2.999997499&6.60E-05&9.58E-05&1&-3.67E-08&29.93779007\\

\hline
3&pareto&2.999828012&0.007848083&0.015613194&0.992&-0.000975087&2.137700346\\

\hline
3&smart5utr&3.000008351&0.00019227&0.000279886&1&-3.13E-07&9.430495812\\

\hline
\end{tabular}\\[10pt]
\end{center}
\end{table*}

\subsubsection{RL, and GC Optimization}

In this subsection, the optimization results for the combination of the RL and the GC optimization features are provided. Table \ref{RL-GC} shows the results of the hybrid optimization, the Pareto optimization, and the Smart5UTR baseline. At RL = 3 and GC = 0.40, and again at RL = 7 and GC = 0.50, the hybrid method effectively achieved zero GC RMSE along with perfect RL success. At the hardest RL = 8 and GC = 0.60 setting, it still held mean RL absolute error to 0.0024 and mean GC absolute error to 0.0239, with 100\% RL-tolerance success. The Smart5UTR baseline also remained accurate in RL, but its GC control deteriorated markedly at RL = 8 and GC = 0.60, where its mean GC absolute error rose to 0.1290. The Pareto method was noticeably weaker on both RL and GC in every RL-GC condition. The mean of the total score for the hybrid method is higher than all of the other methods for all of the target values. This family is one of the clearest demonstrations that explicit multi-condition generation improves controllability beyond a predictor-only design loop.  

\begin{table*}[h]
\small\sf\centering
\begin{center}
\centering
\captionsetup{justification=centering}
\caption{The optimization results for optimizing the combination of target RL and target GC values. \label{RL-GC}}

\begin{tabular}{|c|c|c|c|c|c|c|c|c|c|c|}
\hline
Scenario&Method&Target RL&Target gc&Mean RL&RL MAE&RL Rmse&gc MAE&GC Rmse&Mean total score\\

\hline
RL-GC-RL3-GC040&hybrid&3&0.4&3.000012281&0.000252434&0.000397035&0&0&-6.31E-07\\

\hline
RL-GC-RL3-GC040&Pareto&3&0.4&3.000960096&0.010553719&0.013437754&0.12056&0.134241573&-0.012778293\\

\hline
RL-GC-RL3-GC040&smart5utr&3&0.4&2.999975823&0.000645909&0.001000251&0&0&-4.00E-06\\

\hline
RL-GC-RL7-GC050&hybrid&7&0.5&6.999997767&7.91E-05&0.000117284&0&0&-5.50E-08\\

\hline
RL-GC-RL7-GC050&Pareto&7&0.5&6.95454796&0.050984802&0.091581958&0.023&0.034374409&-0.03584902\\

\hline
RL-GC-RL7-GC050&smart5utr&7&0.5&7.000011076&0.000203735&0.00029313&0&0&-3.44E-07\\

\hline
RL-GC-RL8-GC060&hybrid&8&0.6&7.998594309&0.002434797&0.00490738&0.02392&0.06006663&-0.00248833 \\

\hline
RL-GC-RL8-GC060&Pareto&8&0.6&7.149684685&0.850418884&0.887487065&0.07832&0.096771897&-3.158365161 \\

\hline
RL-GC-RL8-GC060&smart5utr&8&0.6&7.99413897&0.00723614&0.010027388&0.12904&0.151049661&-0.013306194 \\

\hline
\end{tabular}\\[10pt]
\end{center}
\end{table*}  

\subsubsection{RL, and MFE Optimization}
The RL and MFE optimization family shows that the hybrid model learns free-energy control extremely well. Across the three RL-MFE scenarios, the hybrid method achieved mean MFE absolute error of exactly 0.0000 in Table \ref{RL-MFE} while keeping mean RL absolute error between 0.00015 and 0.00139 and maintaining 100\% success. The Smart5UTR baseline was still strong in this family, but not as exact, reaching mean RL absolute error up to 0.0174 and mean MFE absolute error up to 0.0070. The Pareto method again lagged behind, with mean MFE absolute errors of 0.0550 to 0.3898 and RL success rates between 0.8\% and 90.0\%. Since RNA secondary structure and folding energetics are known to influence initiation efficiency, the strong RL-MFE performance of the hybrid model reinforces the manuscript’s design rationale for jointly conditioning on translational and structural variables. \citep{babendure2006control} Similar to the previous subsections, the mean of the total score for the hybrid method is higher than all of the other methods for all of the target values.

\begin{table*}[h]
\small\sf\centering
\begin{center}
\centering
\captionsetup{justification=centering}

\caption{The optimization results for optimizing the combination of target RL and target MFE values.}
\label{RL-MFE}

\resizebox{\textwidth}{!}{%
\begin{tabular}{|c|c|c|c|c|c|c|c|c|c|c|}
\hline
Scenario &
Method &
Target RL &
Target MFE &
Mean RL &
RL MAE &
RL RMSE &
MFE MAE &
MFE RMSE &
RL Success rate &
Mean total score \\
\hline

RL-MFE-RL8-MFE-5 & hybrid & 8 & -5 &
7.999057665 & 0.001388521 & 0.006025196 &
0 & 0 & 1.000 & -0.000145212 \\
\hline

RL-MFE-RL8-MFE-5 & Pareto & 8 & -5 &
7.433502648 & 0.567388667 & 0.592598797 &
0.389799973 & 0.545985343 & 0.008 & -1.599593326 \\
\hline

RL-MFE-RL8-MFE-5 & smart5utr & 8 & -5 &
7.983267112 & 0.017367249 & 0.040785018 &
0.006999993 & 0.026457488 & 0.948 & -0.010153668 \\
\hline

RL-MFE-RL7-MFE-6 & hybrid & 7 & -6 &
6.999993032 & 0.000149922 & 0.000222097 &
0 & 0 & 1.000 & -1.97E-07 \\
\hline

RL-MFE-RL7-MFE-6 & Pareto & 7 & -6 &
6.972313084 & 0.040065308 & 0.058367759 &
0.054999956 & 0.094339754 & 0.900 & -0.041127159 \\
\hline

RL-MFE-RL7-MFE-6 & smart5utr & 7 & -6 &
6.999970808 & 0.000395300 & 0.000621439 &
0 & 0 & 1.000 & -1.54E-06 \\
\hline

RL-MFE-RL3-MFE-10 & hybrid & 3 & -10 &
3.000050041 & 0.000502784 & 0.000713904 &
0 & 0 & 1.000 & -2.04E-06 \\
\hline

RL-MFE-RL3-MFE-10 & Pareto & 3 & -10 &
3.013401179 & 0.055359579 & 0.074045147 &
0.120400120 & 0.191937535 & 0.836 & -0.082130795 \\
\hline

RL-MFE-RL3-MFE-10 & smart5utr & 3 & -10 &
2.999968677 & 0.001305501 & 0.002116359 &
0.000200001 & 0.004472153 & 1.000 & -0.000117916 \\
\hline

\end{tabular}%
}
\end{center}
\end{table*}

\subsubsection{RL, GC, and MFE Optimization}

The single RL+GC+MFE scenario results are provided in table \ref{RL-GC-MFE}. Here the target values for all of the methods are RL = 8, GC = 0.50, and MFE = -10. In this case, the hybrid method still performed best overall: mean optimized RL was 7.974, RL MAE was 0.028, GC MAE was 0.110, MFE MAE was 0.005, and RL-tolerance success rate was 93.0\%. The Smart5UTR baseline was clearly weaker, with RL MAE 0.0943, GC MAE 0.1634, MFE MAE 0.0240, and only 55.0\% RL-tolerance success. The Pareto method nominally had slightly lower GC error than the hybrid model in this one scenario, but its RL MAE ballooned to 0.662, its MFE MAE to 0.7530, and its RL success to only 0.8\%. That pattern is important for the paper: when objectives become jointly restrictive, the proposed method preserves balanced performance, whereas simpler alternatives tend to optimize one property at the expense of the others.

\begin{table*}[h]
\small\sf\centering
\begin{center}
\captionsetup{justification=centering}
\caption{The optimization results for optimizing the combination of the target RL, the target GC, and target MFE values.}
\label{RL-GC-MFE}

\begin{tabular}{|c|c|c|c|c|c|c|c|c|c|c|c|c|}
\hline

Method &
Mean optimized RL &
RL MAE &
RL RMSE &
GC MAE &
GC RMSE &
MFE MAE &
MFE RMSE &
Success RL rate &
Mean total score\\
\hline

Hybrid &
7.974&
0.028&
0.048&
0.110&
0.156&
0.005&
0.022&
0.93&
-0.023\\
\hline

Pareto &
7.337&
0.662&
0.685&
0.104&
0.123&
0.753&
1.088&
0.008&
-2.265\\
\hline

Smart5UTR &
7.906&
0.094&
0.127&
0.163&
0.189&
0.024&
0.054&
0.55&
-0.093\\
\hline

\end{tabular}
\end{center}
\end{table*}

\subsubsection{RL, and structure Optimization}
Structure-constrained design is where the hybrid model most clearly separates itself from the two baseline comparators. The results of the optimization of the RL and structure are provided in Table \ref{RL-Structure}. The target structure for all of the scenarios in Table \ref{RL-Structure} is ``....(((...)))......((....)).......................''.  Averaged across the four RL-structure scenarios, the hybrid method achieved mean RL absolute error 0.0007, mean structure match 0.8752, and 100\% RL success. The Smart5UTR baseline remained RL-accurate, with mean RL absolute error 0.0009, but its mean structure match was lower at 0.8343. The Pareto method reached only 0.7393 mean structure match and just 51.8\% RL success on average. At the highest RL-structure target (RL = 8 with a fixed dot-bracket structure), the hybrid method achieved the structure match 0.8440 versus 0.8260 for the Smart5UTR baseline and 0.7724 for the Pareto method. In a manuscript centered on structure-aware modified-mRNA design, this is one of the strongest result families because it shows that adding explicit structure conditioning improves not only structural fidelity but also overall target stability. The mean of the total score for the hybrid method is again higher than all of the other methods for all of the target values. 

\begin{table*}[h]
\small\sf\centering
\begin{center}
\captionsetup{justification=centering}
\caption{The optimization results for optimizing target RL values together with a fixed target secondary structure.}
\label{RL-Structure}

\begin{tabular}{|c|c|c|c|c|c|c|c|c|c|c|}
\hline
Scenario&
Method &
Target
RL &
Mean optimized RL&
RL MAE&
RL
RMSE &
Mean structure match&
Structure RMSE &
Mean total score\\
\hline

1 &
Hybrid &
8 &
7.999463026 &
0.000794146 &
0.005033263 &
0.8439 &
0.160725854 &
-0.078121335 \\
\hline

1 &
Pareto &
8 &
6.866836871 &
1.133163129 &
1.204485952 &
0.7723 &
0.236587405 &
-5.916965631 \\
\hline

1 &
Smart5UTR &
8 &
7.999774851 &
0.000554583 &
0.00339176 &
0.8260  &
0.17757252 &
-0.087026016 \\
\hline

2 &
Hybrid &
7 &
6.999914319 &
0.000303445 &
0.001644447 &
0.8990 & 
0.106331557 &
-0.050470817 \\
\hline

2 &
Pareto &
7 &
6.774791238 & 
0.229328223 &
0.370941535 &
0.7864 & 
0.217505862 &
-0.657190491 \\
\hline

2 &
Smart5UTR &
7 &
7.000063489 &
0.000381755 &
0.001355016 &
0.86464 &
0.142632395 &
-0.067687344 \\
\hline

3 &
Hybrid &
5 &
5.000098609 &
0.000859534 &
0.002965753 &
0.8951 &
0.10958102 &
-0.052475183 \\
\hline

3 &
Pareto &
5 &
4.999670605 &
0.007601894 &
0.010923258 &
0.8039 &
0.196580772 &
-0.09849727 \\
\hline

3 &
Smart5UTR &
5 &
5.000361017 &
0.001410673 &
0.004779794 &
0.8245 &
0.183808596 &
-0.087811386 \\
\hline

4 &
Hybrid &
3 &
3.000138197 & 
0.000938143 &
0.002117722 &
0.8626 &
0.143390376 &
-0.068717939 \\
\hline

4 &
Pareto &
3 &
3.094325909 &
0.105078498 &
0.140706052 &
0.5942 &
0.422702732 &
-0.282052773 \\
\hline

4 &
Smart5UTR &
3 &
3.000003084 &
0.001260334 &
0.003326127 &
0.8218 &
0.185836487 &
-0.089104252 \\
\hline

\end{tabular}
\end{center}
\end{table*}

\subsubsection{Forbidden motifs experiments 
}

The forbidden motifs scenarios again favored the hybrid model for robust accuracy, especially at the higher RL target. The results of the combination of forbidden motifs and RL is provided in Table \ref{RL-F-motif}. The hybrid method remained exact at both RL = 5 and RL = 7, whereas the Pareto method dropped to mean RL absolute error 0.0759 and 70.0\% success at RL = 7. The forbidden motif success rate, and the mean total score for the hybrid method is higher than the Pareto method and Smart5UTR for all of the forbidden motifs experiments. In addition, the RL RMSE for the hybrid method is lower than the Pareto method and Smart5UTR. Therefore, the hybrid method generated better sequences than the Pareto and Smart5UTR methods. 

\begin{table*}[h]
\small\sf\centering
\begin{center}
\captionsetup{justification=centering}
\caption{The optimization results for optimizing target RL values together with forbidden motifs.}
\label{RL-F-motif}

\begin{tabular}{|c|c|c|c|c|c|c|c|c|}
\hline
Scenario&
Method &
Target
RL &
Mean optimized RL&
RL MAE&
RL
RMSE &
Forbidden motif success rate&
Mean total score  &
Mean runtime (sec)\\
\hline

1 &
Hybrid &
5 &
5.000009741 &
7.60E-05 &
0.000108526 &
1 &
0.249999953 &
29.29870084 \\
\hline

1 &
Pareto &
5 &
5.004004223 &
0.012660034 &
0.022120334 &
0.998 &
0.247292763 &
2.178902365 \\
\hline

1 &
Smart5UTR &
5 &
4.999991375 &
0.000200747 &
0.000314382 &
0.186 &
-0.055250395 &
9.02912359 \\
\hline

2 &
Hybrid &
7 &
6.999998089 &
2.72E-05 &
4.00E-05 &
1 &
0.249999994 &
27.29244194 \\
\hline

2 &
Pareto &
7 &
6.926999156 &
0.075924048 &
0.122257935 &
0.878 &
0.144461989 &
2.406232598  \\
\hline

2 &
Smart5UTR &
7 &
6.999996993 &
7.47E-05 &
0.000108848 &
0.998 &
0.249249953 &
8.79814246 \\
\hline

\end{tabular}
\end{center}
\end{table*}

\subsection{Biologically Motivated Regulatory Element Design}

Although the previous experiments evaluated the ability of the proposed framework to satisfy global design objectives such as ribosome load, GC content, minimum free energy, and secondary structure, practical mRNA engineering also requires precise control over well-established regulatory elements within the 5' untranslated region (5' UTR). These sequence elements regulate translation initiation through distinct molecular mechanisms, including ribosome recognition, leaky scanning, upstream translation, ribosome recruitment, and local RNA accessibility \citep{morris2000upstream, wang20045}. Consequently, successful generation of these motifs provides a more biologically meaningful evaluation than optimization of numerical objectives alone. 

To evaluate this capability, six biologically motivated design scenarios were considered: Kozak sequences, upstream AUGs (uAUGs), upstream open reading frames (uORFs), terminal oligopyrimidine (TOP)-like motifs, translation initiator of short 5' UTR (TISU) elements, and cap-proximal accessibility. The exact sequence constraints used for each experiment are summarized below. 

\subsubsection{Kozak Sequence Generation}
The Kozak sequence is the canonical translation initiation context surrounding the AUG start codon in eukaryotic mRNAs. Experimental studies by Kozak demonstrated that nucleotides immediately flanking the AUG strongly influence ribosome recognition and translation initiation efficiency, with a purine at position -3 and guanine at position +4 providing the optimal initiation context. Since then, Kozak sequences have become a fundamental design principle in recombinant protein expression, gene therapy vectors, and mRNA therapeutics because strengthening the Kozak context generally increases protein production \citep{kozak1986point, kozak1986influences, kozak1987least, kozak1989circumstances}. 

In this work, the required Kozak motif was $\mathbf{(GCC)RCCAUGG}$, where R denotes either adenine or guanine. 

The benchmark evaluated whether generated sequences simultaneously satisfied the target ribosome load while correctly incorporating this Kozak initiation context. 

The results for the Kozak motif optimization with a target RL of 7.5 is provided in Table \ref{Kozak}. The proposed hybrid framework achieved 100\% exact Kozak success, demonstrating that every generated sequence correctly incorporated the required Kozak motif while satisfying the remaining optimization objectives. The Smart5UTR optimization baseline achieved 95.2\% success, whereas the Pareto method achieved only 0.8\%, indicating that the hybrid optimization strategy more effectively preserved both sequence-level regulatory motifs and global translation objectives. 

\textbf{Applications}: Strong Kozak sequences are routinely incorporated into synthetic expression vectors, mRNA vaccines, therapeutic mRNA constructs, and recombinant protein production systems because they substantially improve translation initiation efficiency and protein yield \citep{kozak1989circumstances}.

\begin{table*}[h]
\small\sf\centering
\begin{center}
\captionsetup{justification=centering}
\caption{The optimization results for optimizing target RL values together with the Kozak motif.}
\label{Kozak}

\begin{tabular}{|c|c|c|c|c|c|c|c|c|}
\hline
Method &
Target
RL &
Mean optimized RL&
RL MAE&
RL
RMSE &
Mean total score &
Mean kozak score &
Kozak exact success rate \\
\hline

Hybrid &
7.5 &
7.500001211 &
5.13E-05 &
7.57E-05 &
1.999948677 &
1 &
1  \\
\hline

Pareto &
7.5 &
7.08950288 &
0.412171133 &
0.495461642 &
1.490859248 &
0.7548 &
0.008 \\
\hline

Smart5UTR &
7.5 &
7.500001645 &
0.000135232 &
0.000251103 &
1.995064831 &
0.9952 &
0.952 \\
\hline

\end{tabular}
\end{center}
\end{table*}

\subsubsection{TOP-like Motifs}
Terminal oligopyrimidine (TOP) motifs are pyrimidine-rich sequences located immediately downstream of the 5' cap. Naturally occurring TOP mRNAs primarily encode ribosomal proteins and translation factors, and their translation is tightly regulated through the mTOR signaling pathway \citep{philippe2020global, farooq2022amino}. 

The TOP-like motif enforced in this study consisted of $\mathbf{CY_7}$, that is, an initial cytosine followed by at least seven consecutive pyrimidines (cytosine or uracil). In the implementation, only the first seven nucleotides of the generated 5' UTR are examined. The first nucleotide must be C, and at least 6 of these seven nucleotides must be pyrimidines (C or U). 

The results of the optimization of TOP-like motifs with the target RL of 7.5 is provided in Table \ref{TOP}. The hybrid model achieved 100\% exact success in generating valid TOP-like motifs while preserving the remaining optimization objectives, whereas the Smart5UTR baseline achieved 99.6\% success. These results demonstrate that the proposed framework can accurately integrate extended pyrimidine-rich regulatory regions into generated 5' UTRs. 

\textbf{Applications}: TOP motifs regulate ribosome biogenesis, cellular growth, mTOR-dependent protein synthesis, and synthetic nutrient-responsive expression systems.

\begin{table*}[h]
\small\sf\centering
\begin{center}
\captionsetup{justification=centering}
\caption{The optimization results for optimizing target RL values together with the TOP-like motif.}
\label{TOP}

\begin{tabular}{|c|c|c|c|c|c|c|c|c|}
\hline
Method &
Target
RL &
Mean optimized RL&
RL MAE&
RL
RMSE &
Mean total score &
Mean top-like score &
top-like success rate \\
\hline

Hybrid &
7.5 &
7.500001597 &
4.30E-05 &
6.33E-05 &
1.999957015 &
1 &
1  \\
\hline

Pareto &
7.5 &
6.917643927 &
0.582977369 &
0.638823291 &
1.438986183 &
0.788457143 &
0.22  \\
\hline

Smart5UTR &
7.5 &
7.500005968 &
0.000117697 &
0.000166383 &
1.998339474 &
0.998457143 &
0.996 \\
\hline

\end{tabular}
\end{center}
\end{table*}

\subsubsection{Cap-Proximal Accessibility}
Efficient ribosome scanning requires that the region surrounding the translation initiation site remain structurally accessible. Stable secondary structures immediately downstream of the 5' cap or surrounding the AUG codon can substantially reduce translation initiation efficiency, whereas highly accessible regions facilitate ribosome loading and scanning \citep{mustoe2018pervasive, huang2020quality, dichiacchio2016accessfold}. 

In this benchmark, the optimization procedure was required to maintain high accessibility around the translation initiation region while simultaneously satisfying the remaining design objectives. 

The optimization results for accessibility of the generated sequences are provided in Table \ref{ACC}. Both the hybrid framework and the Smart5UTR baseline achieved 100\% accessibility success, indicating that both optimization strategies successfully controlled local RNA structure without sacrificing translation efficiency or other sequence properties. 

\textbf{Applications}: Controlling local accessibility is a fundamental strategy in mRNA vaccine optimization, synthetic gene design, and therapeutic mRNA engineering because ribosome accessibility directly influences protein expression.

\begin{table*}[h]
\small\sf\centering
\begin{center}
\captionsetup{justification=centering}
\caption{The optimization results for accessibility of the generated sequences.}
\label{ACC}

\begin{tabular}{|c|c|c|c|c|c|c|c|c|}
\hline
Method &
Target
RL &
Mean optimized RL&
RL MAE&
RL
RMSE &
Mean total score &
Mean accessibility score &
Accessibility success rate \\
\hline

Hybrid &
7.5 &
7.50000316 &
8.30E-05 &
0.000144703 &
1.99491698 &
0.995 &
1  \\
\hline

Pareto &
7.5 &
7.056339853 &
0.445421469 &
0.511937279 &
1.542606821 &
0.8285 &
0.984  \\
\hline

Smart5UTR &
7.5 &
7.499990267 &
0.000130335 &
0.000199332 &
1.966036372 &
0.966166667 &
1 \\
\hline

\end{tabular}
\end{center}
\end{table*}

\subsection{Benchmark against RhoDesign}
 To evaluate the proposed framework against an existing RNA design approach, we compared our methods with RhoDesign, a recently proposed structure-to-sequence deep learning model that generates RNA sequences from target three-dimensional structural representations \citep{wong2024deep}. Unlike our framework, RhoDesign was developed primarily for RNA aptamer design and optimizes structural consistency rather than translation-related properties such as ribosome load or GC content. Consequently, this benchmark focuses primarily on structural reconstruction quality while additionally reporting the translation-related objectives optimized only by the proposed methods. In addition, RhoDesign is trained only on canonical RNA sequences, not modified RNA sequences.  

To do benchmarking, all methods generated 200 sequences for the same target secondary structure “(((((((......................)))))))..............”. “(((((((......................)))))))” is the RNA secondary structure of an aptamer used in \cite{wong2024deep}. Since our proposed model is trained on 5’UTR sequences with a fixed length of 50 nucleotides, we added dots to the end of the structure “(((((((......................)))))))” to make the total length as 50 nucleotides. 

For the proposed methods, the optimization additionally targeted a RL of 5.5 and a GC content of 0.85, whereas RhoDesign was evaluated without these additional objectives because they are outside the scope of its original design. The benchmark results of the hybrid, pareto, and RhoDesign methods are provided in Table \ref{Bench}.

\begin{table*}[h]
\small\sf\centering
\begin{center}
\captionsetup{justification=centering}
\caption{The Benchmark results of the hybrid and Pareto methods against RhoDesign.}
\label{Bench}

\begin{tabular}{|c|c|c|c|}
\hline
Metric &
Hybrid &
Pareto &
RhoDesign \\
\hline

Mean Structure match &
0.961 &
0.726 &
0.730 \\
\hline

Mean Structure error &
0.039 &
0.274 &
0.270   \\
\hline

Mean Base-pair distance &
7.52 &
7.51 &
11.12 \\
\hline

Mean Base-pair F1 &
0.432 &
0.000 &
0.310 \\
\hline

GC MAE &
0.054 &
0.015 &
0.336 \\
\hline

RL MAE &
0.0052 &
0.0092 &
0.424  \\
\hline

Runtime (s) &
25.37 &
3.90 &
0.27 \\
\hline

\end{tabular}
\end{center}
\end{table*}

 \subsubsection{Hybrid optimization}
The Hybrid method produced sequences whose predicted ribosome load closely matched the desired target. The average predicted ribosome load was 5.498, corresponding to a mean absolute error of only 0.0052, while every generated sequence satisfied the ribosome-load tolerance criterion (100\% success rate). Although GC content was not matched perfectly (mean GC = 0.796 versus the target of 0.85), the Hybrid optimizer achieved excellent structural agreement with the desired secondary structure. 

As we see in Table \ref{Bench}. The average structure match reached 96.1\%, corresponding to a mean structure error of only 3.9\%. The median structure similarity was 96\%, and one quarter of all generated sequences achieved a perfect structural match. The average base-pair F1 score was 0.432, with an average base-pair distance of 7.52. 

These results indicate that the Hybrid optimization successfully balances translation objectives with structural constraints while maintaining very high structural fidelity. 

\subsubsection{Pareto optimization}
As we see in Table \ref{Bench}, the Pareto method also accurately satisfied the ribosome-load objective, producing an average predicted ribosome load of 5.498 with a mean absolute error of 0.0092. Unlike the Hybrid optimizer, however, Pareto matched the GC-content objective much more closely, achieving an average GC content of 0.842, corresponding to an average error of only 0.0147. 

The improved GC optimization came at the expense of structural accuracy. The average structure match decreased to 72.6\%, with an average structure error of 27.4\%. Likewise, all base-pair precision, recall, and F1 scores were essentially zero because many generated structures failed to preserve the intended base-pairing pattern despite achieving a comparable base-pair distance. 

This behavior illustrates the expected trade-off of Pareto optimization, which maintains multiple competing solutions across several objectives rather than aggressively refining a single objective. 

\subsubsection{Comparison with RhoDesign}

As it is shown in Table \ref{Bench}, RhoDesign achieved an average structure match of 73.0\%, a mean structure error of 26.97\%, and an average base-pair F1 score of 0.310. The average base-pair distance was 11.12, which is substantially larger than the 7.52 obtained by the Hybrid method. 

Compared with RhoDesign, the Hybrid framework:  

\begin{enumerate}
\item[(1)] Improved structure match from 73.0\% to 96.1\%;

\item[(2)] Reduced structure error from 26.97\% to 3.9\%;

\item[(3)] Reduced base-pair distance from 11.12 to 7.52; and
\item[(4)] Increased base-pair F1 from 0.310 to 0.432.;
\end{enumerate}

Interestingly, the Pareto optimizer produced structural accuracy comparable to RhoDesign, with a structure match of 72.6\% versus 73.0\%, indicating that Pareto distributed its optimization effort toward satisfying the multiple competing objectives rather than maximizing structural similarity alone. 

An important distinction between the approaches is that RhoDesign was designed exclusively for RNA structural design and therefore does not optimize ribosome load or GC content. In contrast, both proposed optimization strategies simultaneously optimize translation-related objectives together with structural constraints. Consequently, the Hybrid method demonstrates that high structural fidelity can be achieved without sacrificing translation-oriented design objectives, whereas the Pareto strategy provides a broader set of solutions representing different trade-offs among the multiple optimization objectives.

\section{Conclusions}
In this paper, a novel conditional, structure-aware generative model is proposed which can design modified-mRNA 5' UTRs that satisfy translational and biophysical targets more reliably than a predictor-only optimization baseline. The clearest evidence comes from the ablation study, where all learned generative variants outperformed the Smart5UTR baseline, and from the broader scenario sweeps, where the selected hybrid optimization method remained almost exact on different combinations of target RL, GC-content, MFE, structure, and motifs and most real tasks. 

The strongest scientific interpretation is that explicit structure conditioning materially improves design quality for modified mRNA 5' UTRs. In the ablation table, structure-aware models achieved the best total scores, best structure matching, and the best or near-best RL/GC/MFE errors. In the scenario table, the structure-aware hybrid model clearly dominated the RL-structure family and remained robust when multiple objectives were combined.  

The paper’s core claim is well supported: a hybrid conditional generator for modified-RNA 5' UTRs can jointly control ribosome load, GC content, free energy, and secondary structure, and it does so more effectively than simpler search baselines while remaining flexible enough to incorporate motif and prior-based constraints. The benchmark results for comparing the proposed method with RhoDesign was promising as the structure similarity score for the hybrid method was higher than RhoDesign model. In addition, the errors for target GC and RL values for the hybrid method was lower than RhoDesign model. 

For the future work, new wet-lab validations, and direct biophysical measurement for RL and folding metrics could be performed.



\section*{Author Contributions}

Data Curation, Software, Methodology, Visualization, Writing – Original Draft (NZ);
Conceptualization, Funding Acquisition, Supervision, Writing – Review \& Editing (BJY); 
Methodology, Formal Analysis (NZ, BJY, AM).

\section*{Conflicts of interest}

There are no conflicts to declare.

\section*{Acknowledgements}

For the numerical results of this research we have used the computing resources provided by Texas A\&M High Performance Research Computing (HPRC). The authors acknowledge the use of AI for assistance in language editing, coding, and improving the clarity of the manuscript’s presentation. All conceptual development, technical content, analysis, and conclusions are entirely the work of the authors. 





\bibliographystyle{oup-abbrvnat}
\bibliography{reference}

\end{document}